\documentclass[twocolumn,aps,prl]{revtex4-2}
\usepackage{graphicx,adjustbox,multirow,latexsym,amssymb,amsmath,bbold,paralist,braket}

\newcommand{\cphase}{Fm$\bar{3}$m}

\newcommand{\oI}{Pca2$_{1}$}
\newcommand{\oII}{Pmn2$_{1}$}

\newcommand{\aep}{a$_{\text{epi}}$}
\newcommand{\Ef}{$\mid\varepsilon\mid$}

\begin{document}

\title{Origin of Wake-Up Effect in Hafnia}

\author{Aldo Raeliarijaona}
\email{araeliarijaona@carnegiescience.edu}
\author{R. E. Cohen}
\email{rcohen@carnegiescience.edu}
\affiliation{Extreme Materials Initiative, Earth and Planets Laboratory, Carnegie Institution for Science, 5241 Broad Branch Road NW, Washington, DC 20015, USA}

\date{\today}

\begin{abstract}
We performed density functional theory (DFT) calculations on epitaxially strained hafnia. We demonstrate the stabilization of the ferroelectric (\oI) phase from the antiferroelectric ($Pbcn$) in bulk hafnia in the presence of electric field. We found that the polar (\oI) phase can be efficiently stabilized with an adequate choice of film orientation. We show that for a (010)-oriented Pbcn, the ferroelectric \oI~phase can be reached with a relatively small electric field (\Ef $\geq 150$ KV/m). We thus provide a simple explanation to the experimental observation of polarization enhancement through electric field cycling, or wake-up effect, as a ferroelectric phase transition driven by electric field. We find, in contrast, that stress free pure hafnia does not become ferroelectric for any reasonable electric field. So we explain the wake up effect and stabilization of ferroelectric pure hafnia as coming from a combination of epitaxial strain under applied electric field perpendicular to the film. We find that strain (or doping) primarily destabilizes the baddeleyite structure, so that the antiferroelectric Pbcn and ferroelectric phases can form.
\end{abstract}
                          
\maketitle
Ferroelectrics are used in electronics as dielectrics and nonvolatile random access memories (NVFRAMs)\cite{Scott1989}, transistor elements and switches\cite{BosckeFerroelectricityIH2011,Luo2020,khan2020,Balke2019}, but the drive towards smaller electronic devices is a challenge because of the limiting effect of depolarization field, which is inversely proportional to the material thickness \cite{Dawber2003,Junquera_Ghosez2003}. Surprisingly when studying thin hafnia films, which are of particular interest for electronic applications, since hafnia is compatible with silicon, B\"{o}scke et al. \cite{Boscke2011} found polarization at the nanoscale. This was surprising because the depolarizing field was expected to quench polarization in nanoscale ferroelectrics, and furthermore the polar phases of hafnia were not known to be stable. Indeed, early investigations into the refractory and dielectric properties of hafnia demonstrated that it is isomorphous to zirconia (ZrO$_{2}$)\cite{Adam1959,Ruh1970} with the non-polar, monoclinic baddeleyite ($P2_{1}/c$) structure as its ground state at ambient conditions in bulk.  At ambient pressure hafnia undergoes a cubic (\cphase) to tetragonal phase transition at $T \approx$ 2870K with decreasing temperature, and tetragonal to monoclinic transition at $T \approx$ 2000K \cite{Johnson2019}; under pressure other orthorhombic phases (Pbca within [2 GPa,7.5 GPa]\cite{Haines1997,Desgreniers1999,Ohtaka2001,AlKhatatbeh2010}, cotunnite (Pnma) for P $>$ 15 GPa \cite{Lowther1999,Desgreniers1999,Ohtaka2001,AlKhatatbeh2010,Huan2014}) were stabilised but no polar phases.

Many experimental and theoretical studies have attempted to understand ferroelectricity in nanocrystalline hafnia. The polar orthorhombic structures \oI, and \oII~ are considered most relevant.\cite{Huan2014,Johnson2019} Stabilization of the ferroelectric phases were considered by dopants, surface energy, kinetics,  strain, etc.\cite{Sang2015,Park2017,Materlik2015,Liu2019,Qi2020,Xu2021}. Ref.~\onlinecite{Liu2019} showed that the ferroelectric phases are metastable, but did not find conditions where the ferroelectric phase would form. 

Experimentally, hafnia-based samples are usually exposed to an electric field whose strength is around 3.25-3.5 MV/cm (350 MV/m) cycled 10$^{5}$ times\cite{Zhou2013,Starschich2016,Buragohain2019}. This cycling of electric field, which results in an increase in the electric polarization, is known as the \emph{wake-up effect}. It has been proposed that the  \emph{wake-up effect} is due to defect migration.\cite{Zhou2013,Starschich2016}. Since the wake-up effect and fatigue appears to be two opposite sides of the same coin, namely defects migration within a system, it is desirable to limit the number of electric field cycling to avoid early degradation of the device via polarization pinning for example. Understanding the wake-up effect in hafnia-based systems is therefore of fundamental and technological interest.

We performed density functional theory (DFT) computations using {\sc{Quantum Espresso}} (QE) \cite{QE-2009,QE-2017,QE-exa} on pure hafnia as functions of strain and applied electric field. We used Garrity-Bennett-Rabe-Vanderbilt (GBRV)\cite{GBRV2014} pseudopotentials from the standard solid-state pseudopotential (SSSP)\cite{SSSP2018} with PBEsol \cite{PBEsol2008} exchange-correlation. The plane-wave expansion is truncated using a cutoff energy of E$_\textbf{cutoff}$ =  544 eV, and the Brillouin zone was sampled using an 8$\times$8$\times$8 Monkhorst-Pack grid \cite{MP1976}. We find baddeleyite has the lowest energy, as expected, and the relative differences in energy between the different phases track those of other calculations well (Table~\ref{Table1}). The polar structures  \oI~and \oII~have electric polarization of \textbf{P}$=0.51$ C m$^{-2}$ \textbf{y} (crystallographic axis \textbf{b}) and \textbf{P}$=0.6$ C m$^{-2}$ \textbf{z}  (crystallographic axis \textbf{c}) respectively, where \textbf{y} and \textbf{z} indicate Cartesian directions, computed using the Modern Theory of Polarization.\cite{Vanderbilt2018} 
These values match other computed polarization \cite{Batra2017,Qi2020,Xu2021}.
The computed polarization of \oI~ phase is also comparable to the P = 0.41 C m$^{-2}$ measured in La-doped hafnia\cite{Schenk2019}.
\begin{center}
\begin{table}[ht!]
\centering
\begin{adjustbox}{width=\columnwidth,center}
\begin{tabular}{|c|c|c|c|c|c|c|}
\hline
 \multirow{2}{*}{Phases} & \multicolumn{5}{c|}{Lattice Parameters} & \multirow{2}{*}{$\Delta{E}$ (meV/f.u.)} \\
 \cline{2-6}
 &  a (\AA) & b (\AA) & c (\AA ) &  $\beta$ &  & \\
\hline
\multirow{3}{*}{Fm$\bar{3}$m}& 5.03 &  & &  &  This study & 221.60 \\ 
 & 5.04 &   &  &  & Calc. \cite{Qi2020} & 279.1 \cite{Qi2020} \\
 & 5.08 &  &  &  & Exp. \cite{Shanshoury1970} & --- \\
\hline
\multirow{3}{*}{P4$_{2}$nmc} & 3.56 &  & 5.15 & & This study  & 137.95\\
& 3.56 &  & 5.20 & & Calc \cite{Qi2020} & 166.5 \cite{Qi2020} \\
& 3.58 &  & 5.20 & & Exp. \cite{MacLaren2009} & --- \\
\hline
\multirow{2}{*}{Pbcn } & 4.92 &  5.50 &  5.12 & & This study  & 90.9\\
 & 4.93 & 5.66 & 5.14 & & Calc \cite{SaalMaterials2013,KirklinOpen2015} & 66 \cite{SaalMaterials2013,KirklinOpen2015} \\
\hline
Pbca &\multirow{2}{*}{ 5.03} &  \multirow{2}{*}{9.95} & \multirow{2}{*}{5.20} & \multirow{2}{*}{ } & \multirow{2}{*}{This study}  &  \multirow{2}{*}{46.15}\\
 (Brookite)&  &   &  & &  &  \\
\hline
\multirow{3}{*}{Pbca} & 5.13 &  5.25 & 10.07 & & This study  & 28.14 \\
& 5.00 & 5.17 & 9.90 & &Calc \cite{Liu2019} & 33 \\
( OI ) & 5.25 & 5.09 & 10.07 & &Calc \cite{Batra2017} & 20 \\
& 5.13 & 4.96 & 9.79 & &Exp. \cite{Leger1993} & ---\\
\hline
\multirow{3}{*}{Pmn$2_{1}$} & 3.40 &  5.11 & 3.77 & & This study & 117.83  \\
& 3.42 & 5.16 & 3.76 & & Calc. \cite{Qi2020} & 151.9 \cite{Qi2020} \\
& 3.42 & 5.18 & 3.83 & & Calc. \cite{Huan2014} & 150 \cite{Huan2014} \\
\hline
Pnma & 3.29 &  5.51 & 6.43 & & This study  & 291.77  \\
(Cotunnite) & 3.25 & 5.13 & 6.34 & & Calc. \cite{Lowther1999} & 394 \cite{Lowther1999,Barabash2017} \\
\hline
\multirow{3}{*}{Pca$2_{1}$} & 5.21 &  5.00 & 5.02 &  &This study  & 62.38 \\
  & 5.23 & 5.01 & 5.04 & &Calc. \cite{Qi2020} &  90.5 \cite{Qi2020} \\
  & 5.18 & 4.98 & 5.00 & & Calc. \cite{Liu2019} & 53 \cite{Liu2019}\\
\hline
\multirow{3}{*}{P2$_{1}$/c} & 5.08  &  5.16 & 5.18 & 99.06$^{\circ}$ & This study  &  0 \\
& 5.10& 5.15 & 5.29 & 80.35$^{\circ}$ & Calc \cite{Qi2020} & 0 \\
(Baddeleyite)& 5.05& 5.14 & 5.22 & 99.60$^{\circ}$ & Calc. \cite{Liu2019} & 0 \\
& 5.12& 5.17 & 5.29 & 99.11$^{\circ}$ & Exp. \cite{Ruh1970} & ---\\
\hline
\end{tabular}
\end{adjustbox}
\caption{Stable structures of hafnia from first-principles relaxations. $\Delta{E}$ is the energy of the different phases of hafnia relative to the baddeleyite structure. }
\label{Table1}
\end{table}
\end{center}
Epitaxial films are constrained by the substrate in-plane lattice. We study the relative stability of different hafnia phases as functions of epitaxial strain for different orientations relative to the substrate. We consider a square substrate such as Yttria-stabilized zirconia (YSZ) (\aep = 5.144 \AA). Strain $\eta$ was applied by:
\begin{equation}\label{EqStrain}
a_{\eta} =  (1+\eta) a_{0},
\end{equation}
where $a_{0}$ is the strain-free ground state lattice constant of the tetragonal phase ($a_{epi} = 5.00$ \AA). The unconstrained lattice constants and the atomic positions were relaxed at each strain.

The electric enthalpy is given by:
\begin{equation}\label{ElecH}
\mathcal{F}=U_{KS} - \Omega \textbf{P} \cdot \textbf{E},
\end{equation}
where $U_{KS}$, $\Omega$, \textbf{P} and \textbf{E}, are the internal energy, unit cell volume, electric polarization and electric field respectively\cite{NunesGonze2001,IvoSouza2002,FuBellaiche2002}.
The external electric field was applied to the system by constraining the forces on the atoms\cite{FuBellaiche2002}:
\begin{equation}\label{Efield_Force}
    F_{\kappa,i} = \sum_{j}{Z^{*}_{\kappa,ij} \varepsilon_{j}},
\end{equation}
where $Z^{*}_{\kappa,ij}$ is the Born effective charge, with $\kappa$ being the atomic index, i and j the coordinate indices, and $\varepsilon$ the applied electric field. We considered the Z* of the Pbcn structure computed using density functional perturbation theory (DFPT) with the Quantum Espresso code {\sc ph.x}. The application of the external electric field using the constrained atomic forces, rather than the Berry phase\cite{IvoSouza2002}, leads to tractable calculations but more importantly enabled us to fully relax the geometry and the atomic coordinates.

\begin{figure}[!ht]
\centering
\includegraphics[width=1\columnwidth]{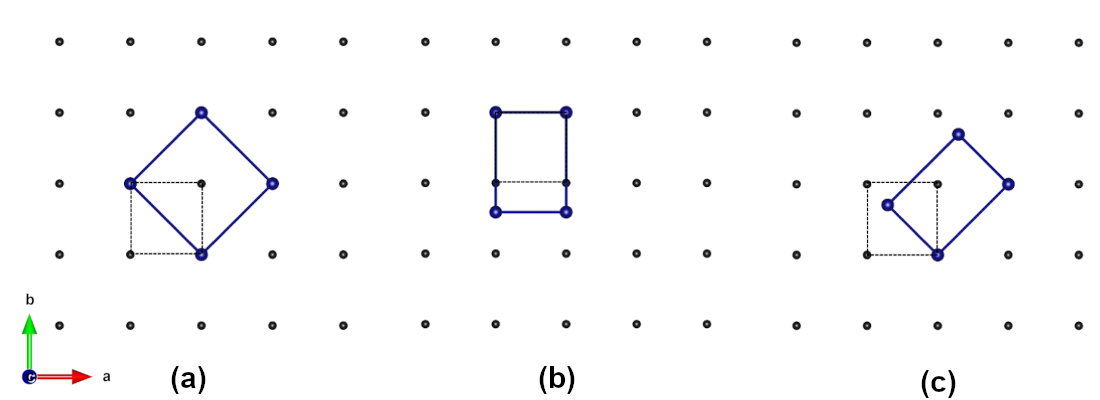}
\caption{Layouts of orthorhombic cells (blue dots delimited by the blue lines) square substrates (black dots delimited by black dotted lines). The orthorhombic cell is laid with (\textbf{a}) the [101] and [$\bar{\text{1}}$01] vectors of the orthorhombic cell matching the substrate diagonals; (\textbf{b}) the [010] vector matching one of the substrate directions with the other in-plane vector ([101] or [$\bar{\text{1}}$01]) allowed to relax; (\textbf{c}) [101] (or the [$\bar{\text{1}}$01] ) vector match the substrate diagonal and the other in-plane vector ([010]) allowed to relax.}
\label{Fig1}
\end{figure}

We considered several ways to lay an orthorhombic cell on a square substrate (Fig.\ref{Fig1}):
\begin{itemize}
\item[i)]{the face diagonals of the orthorhombic cell were matched to the substrate diagonal (Fig.\ref{Fig1}a):
\begin{equation}\label{Eq:aepi_1}
a_{\text{epi}}=\sqrt{\frac{\alpha^{2}+\beta^{2}}{2}};
\end{equation}}
\item[ii)]{one of the lattice vectors match the epitaxial lattice constant (Fig.\ref{Fig1}b) with:
\begin{equation}\label{Eq:aepi_2}
a_{\text{epi}}=\alpha^{2};
\end{equation}}
\item[iii)]{the double of the size of one of the lattice constants is commensurate with the diagonal of the square substrate (Fig.\ref{Fig1}c) for which case we have:
\begin{equation}\label{Eq:aepi_3}
 a_{\text{epi}}=\sqrt{2}\alpha.
\end{equation}}
\end{itemize}
The variables $\alpha$, $\beta$ in Eqs.\ref{Eq:aepi_1}--\ref{Eq:aepi_3} denote the magnitude of the orthorhombic in-plane lattice vectors. When the out-of-plane direction is, for instance, the crystallographic \textbf{c}-direction then $\alpha = a$ and $\beta = b$; when the crystallographic \textbf{b}-direction is out of the epitaxial plane then $\alpha = a$ and $\beta = c$, and if the crystallographic \textbf{a}-direction is out of the epitaxial plane then $\alpha = c$ and $\beta = c$.

In the case of biaxially strained system (Fig.\ref{Fig1}a) the out-of-plane lattice constant was relaxed, whereas for uniaxially strained cases (Fig.\ref{Fig1}b,Fig.\ref{Fig1}c), only the lattice matched, or the diagonally matched lattice constant is fixed, and the remaining axes were relaxed.

\begin{figure*}
\centering
\includegraphics[width=\textwidth]{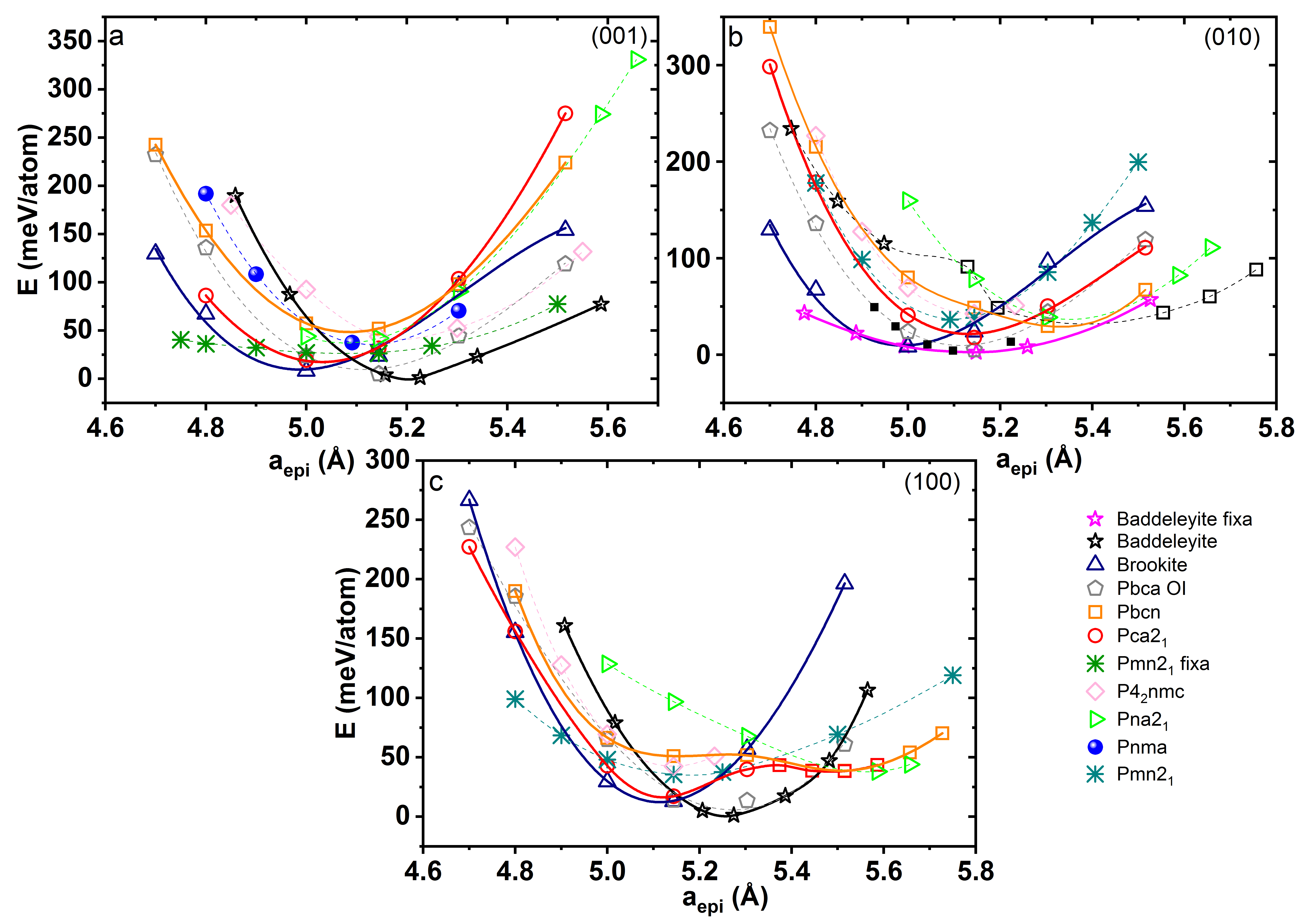}
\caption{Energy vs epitaxial lattice constant in different phases of hafnia. The orientation of the different layouts are such that the crystallographic: (a) \textbf{c}, (b) \textbf{b}, and (c) \textbf{a}-direction is out of the epitaxial plane. This orientation is indicated at the top right corner of each panel. The different phases are plotted with the same color and symbol on the three panels and the uniaxially strained phases, namely baddeleyite and \oII, are drawn with the same symbol as their counterparts but different color (magenta for baddeleyite and green for \oII). The fixed crystallographic axis is for the uniaxially strained systems are noted on the legend. Compressive strain stabilizes the antiferroelectric Pbca structures (Pbca OI and brookite). Under larger compressive strains (\aep$\leq$ 4.9\AA) the uniaxially strained \oII~is most stable.}
\label{Fig2}
\end{figure*}

We find that the (010)-oriented baddeleyite's energy becomes higher than the Pbcn when it is strained biaxially. It even transitions to the Pbcn phase under large epitaxial strains \aep $\geq$ 5.144 \AA (Fig.\ref{Fig2}(b)). 

\begin{figure*}
\centering
\includegraphics[width=\textwidth]{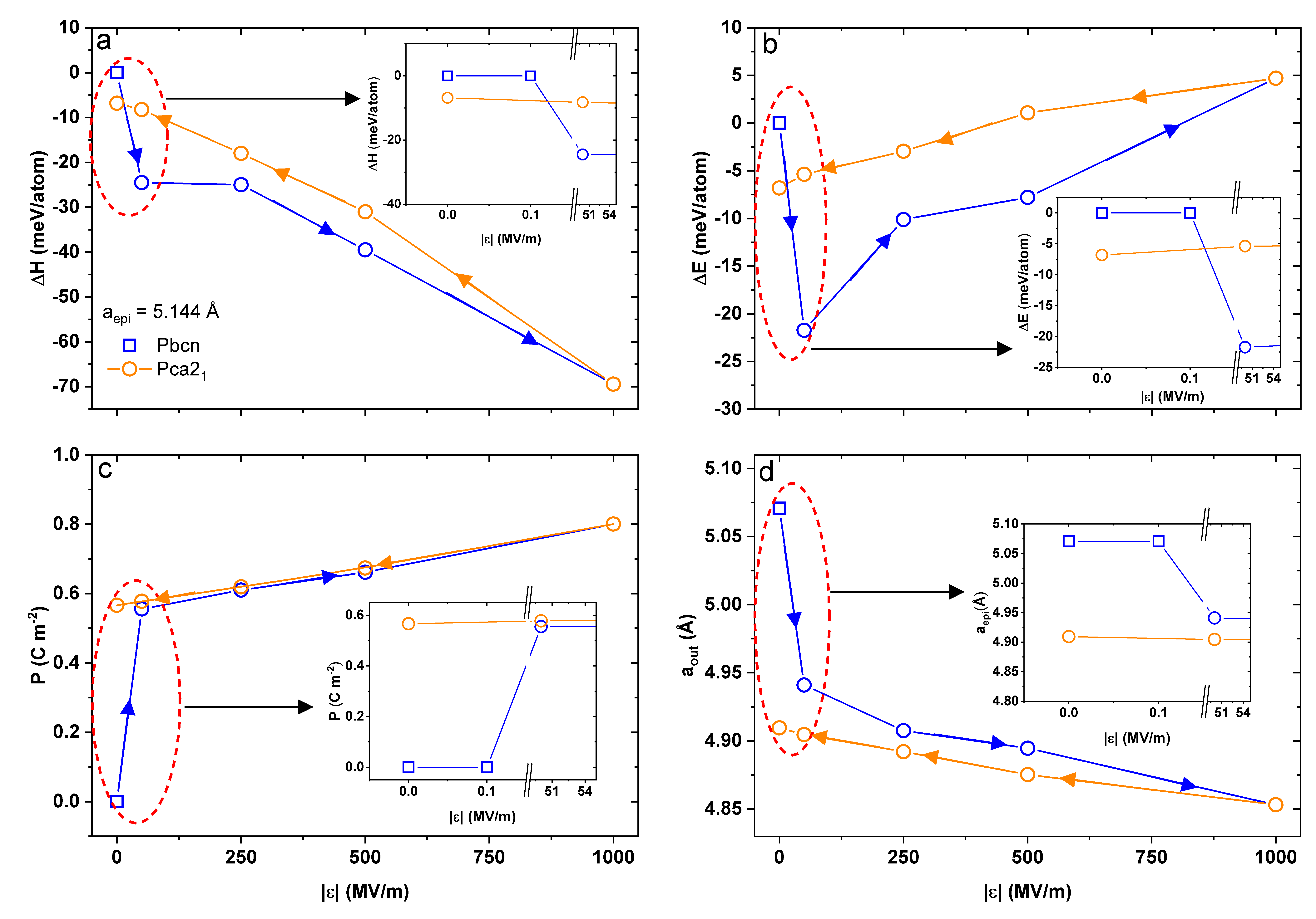}
\caption{Plot of (a) electric enthalpy ($\Delta{\text{H}}$), (b) internal energy ($\Delta{\text{E}}$), (c) electric polarization along the out-of-plane direction, and (d) out-of-plane lattice constant. The inset in each panel is a zoomed-in plot of the region delimited by the dashed line. To avoid repetition, the epitaxial lattice constant and the legend for all four panels are presented in panel (a). The blue (resp. orange) curves were obtained from raising (resp. lowering) the electric field as indicated by the arrows. For the insets, a break in the abscissa was introduced to show the behavior of the curves near the transition. The symmetry at each step was verified using {\sc findsym}\cite{FindSym1,FindSym2}.}
\label{Fig3}
\end{figure*}

We now present the main result of this study, which is the ferroelectric phase transition driven by electric field. Starting from the epitaxially strained (\aep = 5.144 \AA), (010)-oriented Pbcn we applied an electric field normal to the epitaxial plane. The field strength was first raised from 0 to 1000 MV/m, then lowered from 1000 back to 0 MV/m (Fig.\ref{Fig3}). 

We find that at low field values (\Ef $\leq$ 100 KV/m) we are still in the Pbcn phase, but as soon as \Ef $\geq$ 150 KV/m we see a transition from Pbcn to the polar \oI~phase. This \oI~phase stabilizes even more as the field is increased to 1000 MV/m. Decreasing the electric field we find that the system stays in the ferroelectric \oI~phase even at \Ef = 0 MV/m (Fig.\ref{Fig3}(c)). 

(010)-oriented Pbcn has the polarization and applied field perpendicular to the film. The other orientations, (001) and (100) were also considered, which have the polarization and applied fields in plane of the film. In these two orientations, one can still drive the Pbcn hafnia into the ferroelectric \oI~phase, albeit with a stronger field (Supplemental Material\cite{SM}).

Furthermore, we applied electric field to stress-free Pbcn, but no ferroelectric phase transition was observed for a reasonable value of the electric field. For field values \Ef $< 500$ MV/m, the system reverts back to the non-polar Pbcn phase. This finding shows that epitaxial strain is key in the ferroelectric phase transition of hafnia. 

We now discuss its relation to the wake-up effect. The observation of wake-up effect in experiments are usually around 325--350 MV/m \cite{Zhou2013,Starschich2016,Buragohain2019}, which is several orders of magnitude higher than what we found. Though we have transitioned from the non-polar Pbcn to the polar \oI~with a low electric field, the system still has a higher electric enthalpy than the other low energy phases, such as brookite or uniaxially strained baddeleyite, at \Ef $\sim$ 100 KV/m. However, once the electric field values of \Ef $\geq$ 325 MV/m is reached the electric enthalpy of our system becomes the lowest, and as the field is removed hafnia will remains in the \oI~symmetry.

We have shown a second order ferroelectric phase transition driven by an external electric field in hafnia. We have demonstrated further that for the right epitaxial layout and strain, this ferroelectric phase can be reached efficiently (\Ef $\sim$ 150 KV/m) from the Pbcn structure. 

\begin{acknowledgments}
This work is supported by U. S. Office of Naval Research Grant N00014-20-1-2699, and the Carnegie Institution for Science. Computations were supported by  high-performance computer time and resources from the DoD High Performance Computing Modernization Program, Carnegie computational resources, and REC gratefully acknowledges the Gauss Centre for Supercomputing e.V. (https://www.gauss-centre.eu/) for funding this project by providing computing time on the GCS Supercomputer SuperMUC-NG at Leibniz Supercomputing Centre (LRZ, www.lrz.de).  
\end{acknowledgments}

\bibliography{Hafnia}

\end{document}